\documentclass[preprint,prd,noshowpacs, nofootinbib]{revtex4}
\usepackage{amssymb}
\usepackage{amsmath}
\usepackage{graphicx}

\begin{document}

\title{Revisiting the Marton, Simpson, and Suddeth experimental confirmation of the Aharonov-Bohm effect}

\author{James Macdougall}
\email{jbm34@mail.fresnostate.edu}
\author{Douglas Singleton}
\email{dougs@csufresno.edu }
\affiliation{Physics Department, CSU Fresno, Fresno, CA 93740 USA}

\author{Elias C. Vagenas}
\email{elias.vagenas@ku.edu.kw}
\affiliation{Theoretical Physics Group, Department of Physics, Kuwait University,
P.O. Box 5969, Safat 13060, Kuwait}

\date{\today}

\begin{abstract}
We perform an ``archeological" study of one of the original experiments used as  
evidence for the static, {\it time-independent} Aharonov-Bohm effect. Since the experiment in question 
\cite{marton} involved a {\it time varying} magnetic field we show that there are problems with 
the explanation of this experiment as a confirmation of the static Aharonov-Bohm effect -- specifically
the previous analysis ignored the electric field which arises in conjunction with a time-varying
magnetic flux. We further argue that the results of this experiment do in fact conform exactly to the recent 
prediction \cite{singleton, macdougall} of a cancellation between the magnetic and electric 
phase shifts for the {\it time-dependent} Aharonov-Bohm effect. To resolve this issue a new time-dependent 
Aharonov-Bohm experiment is called for.
\end{abstract}

 
\maketitle

The experimental work of Chambers \cite{chambers} is usually cited as the first experimental confirmation 
of the Aharonov-Bohm (AB) effect \cite{AB, siday}. However, the 
earliest (inadvertent) test of the AB effect was an experiment by Marton, Simpson, and Suddeth (MSS) \cite{marton}, 
whose set-up was designed simply to make an electron interferometer to see electron interference fringes, 
and this was apparently accomplished; interference fringes were seen. However, it was realized later that the electrons
in \cite{marton} were moving through a stray 60 Hz magnetic field. Since MSS did see a
fixed interference pattern, it was originally suggested that the effect of this 
magnetic field could be ignored given the 60 keV kinetic energy of the electrons. 
But with the proposal of the AB effect \cite{AB}, it was realized that there should have been a
time-varying shifting of the interference pattern which was not seen in \cite{marton}. Thus, the non-observation
by MSS of a time-varying, shifting interference pattern appeared to be evidence against the
AB effect. However, in a clever paper Werner and Brill \cite{werner} proposed that if one took into account the additional
phase shift due to the direct Lorentz force, that under certain conditions there would be an {\it almost} exact
cancellation of the AB phase shift versus the phase shift coming from, $q(\frac{{\bf v}}{c} \times {\bf B})$. Thus,
according to Werner and Brill, the fact that MSS saw a fixed interference pattern could be taken 
as evidence of the phase shift predicted by the AB effect.

However, the analysis of \cite{werner} ignored the electric field which arises whenever there is a time-varying 
magnetic field. From Faraday's law, {\it i.e.}, $\nabla \times {\bf E} = - \frac{1}{c}\partial _t {\bf B}$, 
one can see that there was an electric field present in the MSS experiment. In terms of potentials, the time-varying magnetic 
field is associated with a time-varying vector potential, namely ${\bf B} ({\bf x}, t) = \nabla \times {\bf A} ( {\bf x}, t)$.
It is this time-varying vector potential that produces an electric field ${\bf E} = - \frac{1}{c}\partial _ t {\bf A} ( {\bf x}, t)$. 
Now,  in \cite{singleton, macdougall} a simple argument was given that the phase shift of this electric field would always 
exactly cancel the AB phase shift coming from the magnetic flux. 
Briefly, the argument of \cite{singleton, macdougall} went as follows: in terms of 
the electric and magnetic fields one can write the AB phase shift, $\alpha$, as
\begin{equation}
\label{phase1}
\alpha = \frac{e}{\hbar} \int _{\it C} F = \frac{e}{\hbar} \left[ \int {\bf E} \cdot d {\bf x} (c dt) + \int {\bf B} \cdot d{\bf S} \right] ~,
\end{equation}
where $F$ is the Faraday two-form which in wedge notation is $(E_x dx + E_y dy + E_z dz ) 
\wedge (c dt) + B_x dy \wedge dz + B_y dz \wedge dx + B_z dx \wedge dy$. 
In the far right-hand side  of Eq. \eqref{phase1} we have reverted to 3-vector notation. Now, for the 
time-dependent Aharonov-Bohm set-up, {\it i.e.}, a solenoid with a time-varying flux, one has a time-varying vector 
potential, ${\bf A} ({\bf x}, t)$, from which one can obtain the fields via ${\bf E} = - \frac{1}{c} \partial_t {\bf A}$ and
${\bf B} = \nabla \times {\bf A}$. As pointed out in \cite{singleton, macdougall}, doing the time integration for 
the electric part of Eq. \eqref{phase1} yields
\begin{equation}
\label{e-phase}
\int {\bf E} \cdot d {\bf x} ~ (c~dt) = - \oint {\bf A} \cdot d {\bf x}  ~,
\end{equation}
and using Stokes' theorem on the magnetic part of Eq. \eqref{phase1} yields
\begin{equation}
\label{b-phase}
\int {\bf B} \cdot d{\bf S} = \oint {\bf A} \cdot d{\bf x} ~.
\end{equation}
Substituting the electric  and magnetic parts, Eq. \eqref{e-phase} and Eq. \eqref{b-phase}, respectively,  in Eq. \eqref{phase1}, 
the total phase shift will be zero, {\it i.e.}, $\alpha =0$. This was the prediction for the time-dependent
AB effect from \cite{singleton, macdougall} and this is in fact what was seen in the MSS experiment. In \cite{singleton}
this cancellation between the was also shown using the path length difference (and therefore phase difference) produced in 
the electron due to the acceleration coming from ${\bf E}$. 

There are several comments to make about the above results: First the above argument only implies a canceling 
of the {\it time-dependent} part of the phase shift {\it not} the time-independent part,{\it i.e.}, the part normally 
referred to as the AB phase shift. If one splits the vector potential in static and time-varying parts as 
${\bf A} ({\bf x}, t) = {\bf A}_0 ({\bf x}) +  {\bf A}_1 ({\bf x}, t)$, it is only the contribution coming from 
${\bf A}_1 ({\bf x}, t)$ which will cancel between Eq. \eqref{e-phase} and Eq. \eqref{b-phase}. In detail 
${\bf B} = \nabla \times {\bf A}_0 ({\bf x}) +  \nabla \times {\bf A}_1 ({\bf x}, t)$ while 
${\bf E} = - \partial_t  {\bf A}_1 ({\bf x}, t)$. Thus it is only the time-dependent part of the vector potential
${\bf A}_1 ({\bf x}, t)$ which takes part in the cancellation between \eqref{e-phase} and \eqref{b-phase} --
the time-independent part ${\bf A}_0 ({\bf x})$ still leads to the usual static AB phase shift. This interplay between 
the electric and magnetic contributions found in Eq. \eqref{e-phase} and Eq. \eqref{b-phase} was noted in a different 
context in Ref. \cite{kampen} (see also \cite{moulopoulos}). In this paper the cancellation between Eq. \eqref{e-phase} and Eq. \eqref{b-phase}
is shown without using the potentials but rather is shown to be a direct consequence of Faraday's Law which in form
notation reads $dF=0$. In detail van Kampen noted that the loop integral of the electric field could be
turned into a surface integral via Stokes' theorem as
\begin{equation}
\label{faraday-3d}
\oint {\bf E} \cdot d{\bf x} = \int \nabla \times {\bf E} \cdot d {\bf S} = 
-\frac{1}{c} \int \partial_t {\bf B}  \cdot d {\bf S} ~,
\end{equation}
where in the last step we have used Faraday's law $\nabla \times {\bf E} = -\frac{1}{c} \partial_t {\bf B}$.
Now applying $\int ... (c ~ dt)$ to \eqref{faraday-3d} we obtain
\begin{equation}
\label{faraday-3d-2}
\int  \left( \oint {\bf E} \cdot d{\bf x} \right) c~dt  = 
- \int \left( \int \partial_t {\bf B}  \cdot d {\bf S} \right) dt = - \int {\bf B}  \cdot d {\bf S} ~,
\end{equation}
where the time integral has been undone via the time integration, and we see that this last
expression in \eqref{faraday-3d-2} cancels the expression in \eqref{b-phase}. Thus following van Kampen
we see that the cancellations is a consequence of the Bianchi identities $d F =0$.  
Second one can ask about the nature of the space-time surfaces in the expression
in \eqref{phase1} versus the surfaces used in the more usually encountered magnetic expression such as in \eqref{b-phase}.
To this end we look at infinitesimal paths and surfaces for the above case and show that the cancellation of the
time-dependent electric and magnetic pieces occurs. For a solenoid oriented along the $z$-axis the vector potential
inside and outside a solenoid, which has a radius $\rho = R$ and a time varying flux, is given by
\begin{equation}
\label{3-vector-a}
{\bf A}_{{\rm in}} = \frac{\rho B(t)}{2} \hat{\bf \varphi} ~~~~{\rm for ~~ \rho <R} ~~~~;~~~~~
{\bf A}_{{\rm out}} = \frac{B(t) R^2}{2 \rho} \hat{\bf \varphi} ~~~~{\rm for ~~ \rho \ge R} ~.
\end{equation}    
The associated magnetic (${\bf B} = \nabla \times {\bf A}$) and electric fields (${\bf E} =\partial _t {\bf A}$) are
\begin{equation}
\label{B-field}
{\bf B}_{{\rm in}} =  B(t) {\bf {\hat z}}~~~~{\rm for ~~ \rho <R} ~~~~;~~~~
{\bf B}_{{\rm out}} = 0  ~~~~{\rm for ~~ \rho \ge R} ~,
\end{equation}
and
\begin{equation}
\label{E-field}
{\bf E}_{{\rm in}} = - \frac{\rho {\dot B(t)}}{2} \hat{\bf \varphi}  
~~~~{\rm for ~~ \rho <R}  ~~~~;~~~~
{\bf E}_{{\rm out}} =  - \frac{{\dot B(t)} R^2}{2 \rho} \hat{\bf \varphi}
~~~~{\rm for ~~ \rho \ge R} ~.
\end{equation}
Now using these expressions in \eqref{B-field} \eqref{E-field} for some short time interval $\Delta t$, 
expanding the magnetic field as $B(t) = B_0 + {\dot B} \Delta t + {\cal O} (\Delta t )^2$, and taking
the path to be a short circular arc segment covered in time $\Delta t$ {\it i.e.}
$\Delta {\bf x} = (\rho \Delta \varphi ) {\hat \varphi}$ we find that the small phase for the
infinitesimal path and associated area are
\begin{equation}
\label{phase2a}
\Delta (\alpha) = \frac{e}{\hbar} \left( {\bf E} \cdot \Delta {\bf x} \Delta t +
\Delta {\bf B} \cdot \Delta {\bf S} \right) = 
\frac{e}{\hbar} \left( - \frac{R^2}{2 \rho} (\dot B \Delta t)(\rho \Delta \varphi ) 
+ \frac{\Delta \varphi R^2}{2} (\dot B \Delta t) \right) = 0 ~.
\end{equation}
Thus one can see the cancellation between the electric and magnetic pieces in \eqref{phase2a} for this 
infinitesimal path and area. Again this cancellation just involves the time-dependent part {\it i.e.}
the $\dot B \Delta t$ part of the magnetic field expansion. The constant first term, $B_0$, would still lead
the time-independent, static AB phase shift. 

We now move on to the re-analysis of the MSS experiment. 
Originally the non-observation of a time shifting interference pattern by MSS led to
the idea that this experiment had ruled out the AB effect. However, in \cite{werner} Werner and Brill 
argued that the non-observation of a time shifting interference pattern (or the fact that MSS
 observed a static interference pattern) was in fact positive evidence for the AB effect due to a 
subtle, {\it almost} cancellation between the magnetic AB phase shift and a phase shift coming from the
$q(\frac{{\bf v}}{c} \times {\bf B})$ force on the electrons. The Werner and Brill explanation of the non-observation 
of a time shifting interference pattern by MSS is conceptually similar to the explanation given 
in \cite{singleton, macdougall} and by Eq. \eqref{e-phase} and Eq. \eqref{b-phase} above. In both cases, there is an interplay
between the magnetic Aharonov-Bohm shifting of the phases and a dispersive shifting of the phases due to either
an electric or magnetic field through which the electrons are moving. However, if one accepts both explanations then 
it seems that one will have over compensated for magnetic AB phase shift and that one should then again see a time 
shifting interference pattern (or the interference pattern would be washed out entirely). We now re-examine the 
Werner and Brill analysis and show that their cancellation is highly dependent on the strength of the unknown 
60 Hz magnetic field in the MSS experiment, and in fact for certain magnetic field strengths 
one does not get the balancing of the AB phase shift against the shift due to the $q(\frac{{\bf v}}{c} \times {\bf B})$ force. 

To start the analysis we examine the re-produced figure 1 from \cite{werner}. In figure 1a, the electron beam is split 
into a diamond pattern by three crystals $C_1$, $C_2$ and $C_3$ in the absence of a magnetic field. In figure 1a, 
the path length $l_1 + l_2$ is equal to the path length of $m_1 + m_2$ so -- modulo the complication that the beams come
together with a small angle $\varphi$ -- one would get constructive interference. Next, in figure 1b, there is a uniform magnetic 
field of magnitude, $B_0$, coming out of the page. This bends the paths $l_1, l_2, m_1, m_2$ into circular arcs of radius
\begin{equation}
\label{radius}
R= \frac{2 \pi \hbar c}{e \lambda B_0} \approx \frac{850}{B_0} {\rm gauss \cdot cm}~.
\end{equation}     
In Eq. \eqref{radius} we use gaussian units and in the last approximation we have plugged in the values of the 
constant $\hbar, c, e$ (in gaussian units) and we have taken the de Broglie wavelength of the electrons from
\cite{marton} as $\lambda = 4.86 \times 10^{-10}$ cm. 

\begin{figure} 
  \centering
	\includegraphics[trim = 0mm 0mm 0mm 0mm, clip, width=5.0cm]{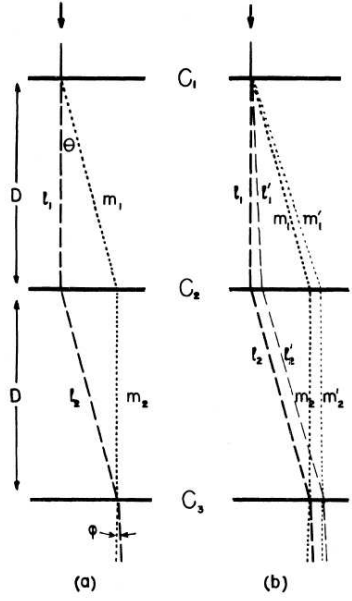}
\caption{The re-produced figure 1 of \cite{werner} showing the paths of the electrons in
the interferometer. Figure 1a is without a magnetic field and figure 1b has a uniform field $B_0$ coming out
of the page. The three diffraction crystals, $C_1$, $C_2$ and $C_3$, are spaced a distance $D$ apart.}
\label{fig1}
\end{figure}

Now, the analysis of \cite{werner} calculated the change in path length produced in the paths $l_1, l_2, m_1, m_2$
due to $B_0$ and then by dividing this path length difference by $\lambda /2 \pi$ obtained a phase difference which when 
combined with the AB phase shift gave an {\it almost} cancellation. The path length differences $\Delta l_2 = l_2 ' - l_2$ and
$\Delta m_1 = m_1 ' - m_1$ were both proportional to 
\begin{equation}
\label{l2m1}
\Delta l_2 \propto \frac{D^2 \tan (\theta)}{R} \;\;\; {\rm and} \;\;\; \Delta m_1 \propto \frac{D^2 \tan (\theta)}{R} ~,
\end{equation}
where from Fig. 1 the angle $\theta$ is the crystal diffraction angle which for the set-up in \cite{marton} was 
$\theta \approx 2 \times 10^{-2}$ radians; $D$ was the distance between the crystals which for the set-up in
\cite{marton} was varied between $5.0$ cm and $3.49$ cm (in subsequent estimates we take $D=5.0$ cm). 
The path length differences between difference $\Delta l_1 = l_1 ' - l_1$ and $\Delta m_2 = m_2 ' - m_2$ 
were ignored since as pointed out \cite{werner} these were of order $B_0 ^2$ whereas the path length 
differences $\Delta l_2$ and $\Delta m_1$ were of order $B_0$. In terms of $D$ and $R$ the path length 
differences $\Delta l_1$ and $\Delta m_2$ can be shown to have a proportionality 
\begin{equation}
\label{l1m2}
\Delta l_1 \propto \frac{D^3}{R^2} = \frac{D^2}{R} \frac{D}{R}\;\;\; {\rm and} \;\;\; 
\Delta m_2 \propto \frac{D^3}{R^2} =  \frac{D^2}{R} \frac{D}{R}~.
\end{equation}
The reason to split off one factor of $D/R$ in Eq. \eqref{l1m2} is to make a comparison between these path length
differences with the path length differences $\Delta m_1$ and $\Delta l_2$ from Eq. \eqref{l2m1}. The ratio
$D/R$ is essentially the angle by which the electrons are bent by the magnetic field, $B_0$. In \cite{werner}
this angle is called $\tau _1$ -- more precisely the angle by which the path $m_1$ is bent is approximated
as $\tau _1 /2$ \cite{werner}. Thus, both the path length differences $\Delta m_1$ and $\Delta l_2$ from
Eq. \eqref{l2m1} and the path length differences  $\Delta m_2$ and $\Delta l_1$ are  of the form
$\frac{D^2}{R} \times angle$, where $angle$ is either the diffraction angle $\tan (\theta) \approx \theta$ 
from Eq. \eqref{l2m1} or the deflection angle $\frac{D}{R}$ due to the magnetic field. The assumption made 
in \cite{werner}, when ignoring the path length differences $\Delta l_1$ and $\Delta m_2$ relative to 
$\Delta l_2$ and $\Delta m_1$, is that $\frac{D}{R} \ll \theta$. Now, from \cite{marton} the diffraction angle
was $\theta \approx 2 \times 10^{-2}$ radians and we take $D \approx 5.0$ cm. Now, even for a 
weak magnetic field of $B_0 = 1$ gauss (which is roughly only twice the Earth's magnetic field 
strength) Eq. \eqref{radius} gives $R \approx 850$ cm and thus $\frac{D}{R} \approx 6 \times 10^{-3}$ radians.
Even for such a weak field the approximation $\frac{D}{R} \ll \theta$ does not really hold (it is true
$\frac{D}{R} < \theta$ but the qualification ``much less than" is not correct). For a magnetic field only 10 times
stronger than the Earth's magnetic field, {\it i.e.}, $B_0 = 5$ gauss , \eqref{radius} gives $R \approx 170$ cm and 
thus $\frac{D}{R} \approx 3 \times 10^{-2}$ radians for which the neither the relationship 
$\frac{D}{R} \ll \theta$ nor even $\frac{D}{R} < \theta$ is valid. 
Thus, depending on the strength of $B_0$ (which was small, but unspecified in the MSS 
experiment) the approximations used by Werner and Brill to ignore $\Delta l_1$ and $\Delta m_2$
might or might not be valid. For magnetic field greater than 1 gauss the approximation becomes quickly suspect. 
The result is that if one has a magnetic field even as weak as $B_0 \approx 2$ gauss one should include 
$\Delta m_2$ and $\Delta l_1$ in calculating the total path length difference which is then
$(\Delta l_2 + \Delta l_1) - (\Delta m_2 + \Delta m_1)$. Moreover, if the diffraction angle, $\theta$, is similar to 
the angle of deviation of the magnetic field, $\frac{D}{R}$, one finds that $\Delta l_1 \sim \Delta m_1 \sim
\Delta l_2 \sim \Delta m_2$ so that there is only a small total path length difference coming from the 
$q(\frac{{\bf v}}{c} \times {\bf B})$ force. In this case the cancellation of the electric phase shift
Eq. \eqref{e-phase} with the usual magnetic phase shift Eq. \eqref{b-phase} is crucial to explain the
results of the MSS experiment. In any case ignoring the effect of the electric field for
the time varying field is not valid while ignoring the effect of $q(\frac{{\bf v}}{c} \times {\bf B})$ is
justifiable depending on the strength of the unknown $B_0$.  

In this paper, we have highlighted problems with previous explanations of and given an alternative explanation for the 
results of the MSS experiment \cite{marton}, which was intended to study electron 
interference, and in fact an interference pattern was observed. However, later it was determined that the
path of the electrons was contaminated with a 60 Hz ``weak" magnetic field, but how weak
was not specified. In light of the predictions for a quantum mechanical, AB phase shift \cite{AB, siday}
which would have naively predicted a shifting/time-varying interference pattern in time to the temporal 
variation of the flux, the observations of a static interference pattern in \cite{marton} seemed, at first,
to provide evidence against the AB phase shift. However, later experiments \cite{chambers} 
with static magnetic fields from iron whiskers did confirm the AB phase shift. The results of the MSS
 experiment were explained in \cite{werner} as an {\it almost} cancellation between the AB phase shift
and a phase shift coming from a path length difference produced by the $q(\frac{{\bf v}}{c} \times {\bf B})$ force
on the electrons since, unlike the ideal AB set-up, the electrons in the MSS experiment
did move in a region with non-zero fields -- both magnetic and electric. This was our first criticism of 
the previous analysis of the MSS experiment and all other studies of the {\it time-dependent}
AB effect -- the effect of the electric field, which arises from the time varying magnetic flux through Faraday's
law, is completely ignored. Further, from our previous works \cite{singleton, macdougall} and the calculations
around Eqs. \eqref{e-phase} and  \eqref{b-phase} above, we find that the phase shift coming from the electric field should 
{\it exactly} cancel the usual magnetic AB phase shift, without any need for the complicated geometrical 
constructions of \cite{werner}. Our second criticism is, depending on the strength of the ``weak"
magnetic field (which was not specified in either \cite{marton} or \cite{werner}) the approximation was made
that the total path length difference was given by $\Delta l_2 - \Delta m_1$ -- thus ignoring the contributions of 
$\Delta l_1$ and $\Delta m_2$ to the total path length difference. This amounts to saying that the diffraction angle, $\theta$,
was much larger than the angle of derivation due to the magnetic field, $\frac{D}{R}$, {\it i.e.},
$\frac{D}{R} \ll \theta$. Using $D \approx 5.0$ cm, $\theta =2 \times 10^{-2}$ radians \cite{marton}
and using Eq. \eqref{radius} for $R$, we showed the even for $B_0$ of a few gauss that the approximation 
$\frac{D}{R} \ll \theta$ was not valid so that one should include $\Delta l_1$ and $\Delta m_2$ in the total
path length difference. Including $\Delta l_1$ and $\Delta m_2$ would decrease the total path length difference 
since for  $\frac{D}{R} \approx \theta$ one has $\Delta l_1 \sim \Delta m_1 \sim \Delta l_2 \sim \Delta m_2$, 
which then gives only a small total path length  difference. This would leave the proposed cancellation of the 
electric and magnetic phase shifts discussed in \cite{singleton, macdougall} as the only explanation of the MSS experiment. 
Finally, the MSS experiment should be considered as an inadvertent test of the 
{\it time-dependent} AB effect rather than as a test of the {\it time-independent} AB effect. 
Here we are proposing that one should re-do a purposeful version of the {\it time-dependent} AB effect experiment 
to determine whether there is really a cancellation between the electric and magnetic phase as suggested in 
\cite{singleton, macdougall}, or if something like the resolution given in \cite{werner} is correct.

As a final comment we know of only one experiment which purposefully sought to test the 
{\it time-dependent} AB effect \cite{chentsov, ageev}, and this experiment did
find no time varying shifting of the interference pattern in agreement with the prediction of 
\cite{singleton, macdougall}. However the authors of \cite{chentsov, ageev} later reported problems 
with the experimental set-up. This again calls for further experiments designed to specifically test the 
{\it time-dependent} AB effect to resolve this issue.


\begin{thebibliography}{99}

\bibitem{marton} L. Marton, J. A. Simpson, and J. A. Suddeth, Rev. Sci. Instr. {\bf 25}, 1099 (1954).

\bibitem{singleton} D. Singleton and E. Vagenas, Phys. Lett. B {\bf 723}, 241 (2013).

\bibitem{macdougall} J. MacDougall and D. Singleton, J. Math. Phys. {\bf 55}, 042101 (2014).

\bibitem{chambers} R.G. Chambers, Phys. Rev. Lett. {\bf 5}, 3 (1960).

\bibitem{AB} Y. Aharonov and D. Bohm, Phys. Rev. {\bf 115}, 484 (1959).

\bibitem{siday} W. Ehrenberg and R. E. Siday, Proc. Phys. Society B {\bf 62}, 8 (1949).

\bibitem{werner} F.G. Werner and D. Brill, Phys. Rev. Lett. {\bf 4}, 344 (1960).

\bibitem{kampen} N.G. van Kampen, Phys. Lett.\ A  {\bf 106}, 5 (1984).

\bibitem{moulopoulos}  K. Moulopoulos, J.\ Phys.\ A  {\bf 43}, 354019 (2010).

\bibitem{chentsov} Yu. V. Chentsov, Yu. M. Voronin, I. P. Demenchonok,
and A. N. Ageev, Opt. Zh. {\bf 8}, 55 (1996).

\bibitem{ageev} A. N. Ageev, S. Yu. Davydov, and A. G. Chirkov, 
Technical Phys. Lett. {\bf 26}, 392 (2000). 

\end{thebibliography}
\end{document}